\input harvmac
\overfullrule=0pt
%
\def\simge{\mathrel{%
   \rlap{\raise 0.511ex \hbox{$>$}}{\lower 0.511ex \hbox{$\sim$}}}}
\def\simle{\mathrel{
   \rlap{\raise 0.511ex \hbox{$<$}}{\lower 0.511ex \hbox{$\sim$}}}}
 
\def\slashchar#1{\setbox0=\hbox{$#1$}           
   \dimen0=\wd0                                 
   \setbox1=\hbox{/} \dimen1=\wd1               
   \ifdim\dimen0>\dimen1                        
      \rlap{\hbox to \dimen0{\hfil/\hfil}}      
      #1                                        
   \else                                        
      \rlap{\hbox to \dimen1{\hfil$#1$\hfil}}   
      /                                         
   \fi}                                         %
\def\ts{\thinspace}
\def\tx{\textstyle}
\def\ra{\rightarrow}

\def\ol{\bar}

\def\CA{{\cal A}}
\def\CB{{\cal B}}
\def\CC{{\cal C}}
\def\CD{{\cal D}}

\def\CF{{\cal F}}

\def\CI{{\cal I}}

\def\CL{{\cal L}}
\def\CM{{\cal M}}
\def\CN{{\cal N}}
\def\CO{{\cal O}}

\def\CS{{\cal S}}

\def\ecm{\sqrt{s}}
\def\shat{\hat s}
\def\that{\hat t}
\def\uhat{\hat u}
\def\rshat{\sqrt{\shat}}

\def\atc{\alpha_{TC}}
\def\aqcd{\alpha_{S}}
\def\atro{\alpha_{\rho_T}}

\def\Ntc{N_{TC}}
\def\suc{SU(3)}

\def\sutc{SU(\Ntc)}

\def\thw{\theta_W}
\def\kslash{\raise.15ex\hbox{/}\kern-.57em k}
\def\LTC{\Lambda_{TC}}
\def\LETC{\Lambda_{ETC}}

\def\CDgg{\CD_{g g}}
\def\CDgrho{\CD_{g\rho_T}}
\def\tro{\rho_{T1}}

\def\troct{\rho_{T8}} 
\def\tropm{\rho_{T1}^\pm}

\def\troz{\rho_{T1}^0}
\def\tpi{\pi_T}
\def\tpipm{\pi_T^\pm}
\def\tpimp{\pi_T^\mp}
\def\tpip{\pi_T^+}
\def\tpim{\pi_T^-}
\def\tpiz{\pi_T^0}
\def\tpipr{\pi_T^{0 \prime}}
\def\etat{\eta_T}
\def\octpi{\pi_{T8}}
\def\octpipm{\pi_{T8}^\pm}
\def\octpip{\pi_{T8}^+}
\def\octpim{\pi_{T8}^-}
\def\octpiz{\pi_{T8}^0}
\def\toppi{\pi_t}
\def\toppip{\pi_t^+}
\def\toppim{\pi_t^-}
\def\toppipm{\pi_t^\pm}
\def\toppiz{\pi_t^0}

\def\Mv{M_{V_8}}
\def\Mzp{M_{Z'}}
\def\tpilq{\pi_{L \ol Q}}

\def\tpiql{\pi_{Q \ol L}}
\def\tpiun{\pi_{U \ol N}}
\def\tpiue{\pi_{U \ol E}}
\def\tpidn{\pi_{D \ol N}}
\def\tpide{\pi_{D \ol E}}

\def\jet{\rm jet}
\def\jets{\rm jets}

\def\pbarp{\ol p p}

\def\gev{{\rm GeV}}
\def\tev{{\rm TeV}}

\def\pb{{\rm pb}}

\def\fb{{\rm fb}}
\def\half{\textstyle{ { 1\over { 2 } }}}

\def\twothirds{\textstyle{ { 2\over { 3 } }}}

\def\myfoot#1#2{{\baselineskip=14.4pt plus 0.3pt\footnote{#1}{#2}}}

\Title{\vbox{\baselineskip12pt\hbox{BUHEP--96--8}
\hbox{hep-ph/9605257}
}}
{Electroweak and Flavor Dynamics at Hadron Colliders}

\bigskip
\centerline{Kenneth Lane\myfoot{$^{\dag }$}{lane@buphyc.bu.edu}}
\smallskip\centerline{Department of Physics, Boston University}
\centerline{590 Commonwealth Avenue, Boston, MA 02215}
\vskip .3in

\centerline{\bf Abstract}

We catalog the principal signatures of electroweak and flavor dynamics at
$\pbarp$ and $pp$ colliders for use at the 1996 Snowmass Workshop on New
Directions in High Energy Physics. The framework for dynamical symmetry
breaking we assume is technicolor, with a walking coupling $\atc$, and
extended technicolor. The reactions discussed occur mainly at subprocess
energies $\rshat \simle 1\,\tev$. They include production of color-singlet
and octet technirhos and their decay into pairs of technipions,
longitudinal weak bosons, or jets. Technipions, in turn, decay
predominantly into heavy fermions. Many of these signatures are also
expected to occur in topcolor-assisted technicolor.  Several particles
specific to this new scenario are discussed. Additional signatures of
flavor dynamics, associated with quark and lepton substructure, may be
sought in excess production rates for high $E_T$ and invariant mass dijets
and dileptons. An important feature of these processes is that they exhibit
fairly central angular and rapidity distributions.

\bigskip

\Date{3/96}

\vfil\eject

\newsec{Plan}

This document lists the major signals for dynamical electroweak and flavor
symmetry breaking in experiments at the Tevatron Collider and the Large
Hadron Collider. It was prepared to help guide studies at the 1996 Snowmass
Summer Study. The motivations for these studies are clear: We do not know
the mechanism of electroweak symmetry breaking nor the physics underlying
flavor and its symmetry breaking. The dynamical scenarios whose signals we
catalog provide an attractive theoretical alternative to perturbative
supersymmetry models. At the same time, they give experimentalists a set of
high-$p_T$ signatures that challenge heavy-flavor tagging, tracking and
calorimetry---detector subsystems somewhat complementary to those tested by
supersymmetry searches. Finally, many of the most important signs of
electroweak and flavor dynamics have sizable rates and are relatively
easily detected in hadron collider experiments. Extensive searches are
underway in both Tevatron Collider collaborations, CDF and D\O. We hope
that this document will help the ATLAS and CMS Collaborations begin their
studies.

Section~2 contains a brief overview of technicolor and extended
technicolor, the best theoretical basis we have for dynamical electroweak
and flavor symmetry breaking. This discussion includes summaries of the
main ideas that have developed over the past decade: walking technicolor,
multiscale technicolor, and topcolor-assisted technicolor.

Hadron collider signals of technicolor involve production of technipions
via $\ol q q$ annihilation and $gg$ fusion. These technipions include the
longitudinal weak bosons $W_L$ and $Z_L$ as well as the pseudo-Goldstone
bosons $\tpi$ of dynamical symmetry breaking. The $\tpi$ are generally
expected to have Higgs-boson-like couplings to fermions and, therefore, to
decay to heavy, long-lived quarks and leptons. The subprocess production
cross sections for color-singlet technipions are listed for simple models
in Section~3.1. The most promising processes involve production of an
isovector technirho $\tro$ resonance and its subsequent decay into
technipion pairs. The most important subprocesses for colored technihadrons
are discussed in Section~3.2. These involve a color-octet $s$-channel
resonance with the same quantum numbers as the gluon; this technirho
$\troct$ dominates colored technipion pair production. It is possible that
$M_{\troct} < 2 M_{\tpi}$, in which case $\troct \ra \ol q q$, $gg$,
appearing as a resonance in dijet production. The main signatures of
topcolor-assisted technicolor, top-pions $\pi_t$ and the color-octet
$V_8$ and singlet $Z'$ of broken topcolor gauge symmetries, are described
in section~3.3.

In Section~4, we motivate and discuss the main ``low-energy'' signatures of
quark and lepton substructure---excess production of high-$E_T$ jets and
high invariant mass dileptons. Cross sections are presented for a simple
form of the contact interaction induced by substructure. We re-emphasize
that the shapes of angular distributions are an important test for new
physics as the origin of such excesses. We also stress the need to study
the effect of other forms for the contact interactions.

This is not intended to be a complete survey of electroweak and flavor
dynamics signatures accessible at hadron colliders. We have limited our
discussion to processes with the largest production cross sections and most
promising signal-to-background ratios. Studies of these processes at
Snowmass will go far toward building a cadre of experts to carry out the
most far-ranging simulations of these processes and their observability in
the detectors now being designed and built. Even for the processes we list,
we have not provided detailed cross sections for signals and backgrounds.
Signal rates depend on masses and model parameters; they and the
backgrounds also depend strongly on detector capabilities. Experimenters in
the detector collaborations will have to carry out these studies. At the
end of this document, I have provided a table summarizing the main
processes, sample cross sections at the Tevatron and LHC, and the names of
CDF and D\O\ members who have experience in these searches and have
graciously agreed to provide guidance for the simulations at Snowmass.


\newsec{Technicolor and Extended Technicolor}

Technicolor---a strong interaction of fermions and gauge bosons at the
scale $\LTC \sim 1\,\tev$---is a scenario for the dynamical breakdown of
electroweak symmetry to electromagnetism
\ref\tcref{S.~Weinberg, Phys.~Rev.~{\bf D19}, 1277 (1979)\semi L.~Susskind,
Phys.~Rev.~{\bf D20}, 2619 (1979).}.
Based on the similar phenomenon of chiral symmetry breakdown in QCD,
technicolor is explicitly defined and completely natural. To account for
the masses of quarks, leptons, and Goldstone ``technipions'' in such a
scheme, technicolor, ordinary color, and flavor symmetries are embedded in
a larger gauge group, called extended technicolor (ETC)
\ref\etc{S.~Dimopoulos and L.~Susskind, Nucl.~Phys.~{\bf B155}, 237
(1979)\semi E.~Eichten and K.~Lane, Phys.~Lett.~{\bf 90B}, 125 (1980).}.
The ETC symmetry is broken down to technicolor and color at a scale $\LETC
= \CO(100\,\tev)$. Technicolor with extended technicolor constitute a
scenario for electroweak and flavor symmetry  breakdown that does not rely
on mystical incantations about physics in hidden sectors at inaccessibly
high energy scales. Indeed, as we describe below, many signatures of ETC
are expected in the energy regime of 100~GeV to 1~TeV, the region covered
by the Tevatron and Large Hadron Colliders. For a review of technicolor
developments up through 1993, see Ref.~\ref\tasi{K.~Lane, {\it An
Introduction  to Technicolor}, Lectures given at the 1993 Theoretical
Advanced Studies Institute, University of Colorado, Boulder, published in
``The Building Blocks of Creation'', edited by S.~Raby and T.~Walker,
p.~381, World Scientific (1994).}.

The principal signals in hadron collider experiments of ``classical''
technicolor and extended technicolor were discussed in
Ref.~\ref\ehlq{E.~Eichten, I.~Hinchliffe, K.~Lane and C.~Quigg,
Rev.~Mod.~Phys.~{\bf 56}, 579 (1984); Phys.~Rev.~{\bf D34}, 1547 (1986).}.
In the minimal technicolor model, containing just one technifermion
doublet, the only prominent signals in high energy collider experiments are
the modest enhancements in longitudinally-polarized weak boson production.
These are the $s$-channel color-singlet technirho resonances near
1.5--2~TeV: $\troz \ra W_L^+W_L^-$ and $\tropm \ra W_L^\pm Z_L^0$. The
small $O(\alpha^2)$ cross sections of these processes and the difficulty of
reconstructing weak-boson pairs with reasonable efficiency make observing
these enhancements a challenge. Nonminimal technicolor models are much more
accessible because they have a rich spectrum of lower energy technirho
vector mesons and technipion ($\tpi$) states into which they may decay. In
the one-family model, containing one isodoublet each of color-triplet
techniquarks $(U,D)$ and color-singlet technileptons $(N,E)$, the
technifermion chiral symmetry is $SU(8) \otimes SU(8)$. There are~63
$\rho_T$ and $\tpi$, classified according to how they transform under
ordinary color $SU(3)$ times weak isospin $SU(2)$. The technipions are
$\tpipr \in (1,1)$; $W^\pm_L, Z^0_L$ and $\tpipm, \tpiz \in (1,3)$;  color
octets $\etat \in (8,1)$ and $\octpipm, \octpiz \in (8,3)$; and
color-triplet leptoquarks $\tpiql,\ts \tpilq \in (3,3) \oplus (3,1) \oplus
(\ol 3,3)\oplus(\ol 3,1)$. The $\rho_T$ belong to the same representations.

Because of the conflict between constraints on flavor-changing neutral
currents and the magnitude of ETC-generated quark, lepton and technipion
masses, classical technicolor was superseded a decade ago by ``walking''
technicolor. In this kind of gauge theory, the strong technicolor coupling
$\atc$ runs very slowly for a large range of momenta, possibly all the way
up to the ETC scale---which must be several 100~TeV to suppress FCNC. This
slowly-running coupling permits quark and lepton masses as large as
a few~GeV to be generated from ETC interactions at this very high scale
\ref\wtc{B.~Holdom, Phys.~Rev.~{\bf D24}, 1441 (1981);
Phys.~Lett.~{\bf 150B}, 301 (1985)\semi
T.~Appelquist, D.~Karabali and L.~C.~R. Wijewardhana,
Phys.~Rev.~Lett.~{\bf 57}, 957 (1986);
T.~Appelquist and L.~C.~R.~Wijewardhana, Phys.~Rev.~{\bf D36}, 568
(1987)\semi 
K.~Yamawaki, M.~Bando and K.~Matumoto, Phys.~Rev.~Lett.~{\bf 56}, 1335
(1986) \semi
T.~Akiba and T.~Yanagida, Phys.~Lett.~{\bf 169B}, 432 (1986).}.

Walking technicolor models require a large number of technifermions in
order that $\atc$ runs slowly. These fermions may belong to many copies of
the fundamental representation of the technicolor gauge group, to a few
higher dimensional representations, or to both. This fact inspired a new
kind of model, ``multiscale technicolor'', and a very different
phenomenology
\ref\multi{K. Lane and E. Eichten, Phys. Lett. {\bf B222}, 274 (1989)\semi
K.~Lane and M.~V.~Ramana, Phys.~Rev.~{\bf D44}, 2678 (1991).}.
In multiscale models, there typically are two widely separated scales of
electroweak symmetry breaking, with the upper scale set by the weak decay
constant $F_\pi = 246\,\gev$. Technihadrons associated with the lower scale
may be so light that they are within reach of the Tevatron collider; they
certainly are readily produced {\it and detected} at the LHC. Because of
technipion mass enhancements in walking technicolor models, some $\rho_T
\ra \tpi\tpi$ decay channels may be closed. If this happens with
color-octet $\troct$, these copiously produced states appear as resonances
in dijet production. If the $\tpi\tpi$ channels are open, they are
resonantly produced at large rates---of order 10~pb at the Tevatron and
several nanobarns at the LHC---and, given the recent successes and coming
advances in heavy flavor detection, many of these technipions should be
reconstructable in the hadron collider environment.

Another major advance in technicolor came in the past two years with the
discovery of the top quark
\ref\toprefs{F.~Abe, et al., The CDF Collaboration, Phys.~Rev.~Lett.~{\bf
73}, 225 (1994); Phys.~Rev.~{\bf D50}, 2966 (1994); Phys.~Rev.~Lett.~{\bf
74}, 2626 (1995) \semi
S.~Abachi, et al., The D\O\ Collaboration, Phys.~Rev.~Lett.~{\bf
74}, 2632 (1995).}.
Theorists have concluded that ETC models cannot explain the top quark's
large mass without running afoul of either cherished notions of naturalness
or experimental constraints from the $\rho$ parameter and the $Z \ra \ol b
b$ decay rate
\ref\zbbexp{P.~B.~Renton, Rapporteur talk at the International Conference
on High Energy Physics, Beijing (August 1995);
LEP Electroweak Working Group, LEPEWWG/95-02 (August 1, 1995).},
\ref\zbbth{R.~S.~Chivukula, S.~B.~Selipsky, and E.~H.~Simmons,
Phys.~Rev.~Lett.~{\bf 69} 575, (1992)\semi
R.~S.~Chivukula, E.~H.~Simmons, and J.~Terning,
Phys.~Lett.~{\bf B331} 383, (1994), and references therein.}.
This state of affairs has led to ``topcolor-assisted technicolor'' (TC2).
In TC2, as in top-condensate models of electroweak symmetry breaking
\ref\topcondref{Y.~Nambu, in {\it New Theories in Physics}, Proceedings of
the XI International Symposium on Elementary Particle Physics, Kazimierz,
Poland, 1988, edited by Z.~Adjuk, S.~Pokorski and A.~Trautmann (World
Scientific, Singapore, 1989); Enrico Fermi Institute Report EFI~89-08
(unpublished)\semi
V.~A.~Miransky, M.~Tanabashi and K.~Yamawaki, Phys.~Lett.~{\bf
221B}, 177 (1989); Mod.~Phys.~Lett.~{\bf A4}, 1043 (1989)\semi
W.~A.~Bardeen, C.~T.~Hill and M.~Lindner, Phys.~Rev.~{\bf D41},
1647 (1990).},
\ref\topcref{C.~T. Hill, Phys.~Lett.~{\bf 266B}, 419 (1991) \semi
S.~P.~Martin, Phys.~Rev.~{\bf D45}, 4283 (1992);
{\it ibid}~{\bf D46}, 2197 (1992); Nucl.~Phys.~{\bf B398}, 359 (1993);
M.~Lindner and D.~Ross, Nucl.~Phys.~{\bf  B370}, 30 (1992)\semi
R.~B\"{o}nisch, Phys.~Lett.~{\bf 268B}, 394 (1991)\semi
C.~T.~Hill, D.~Kennedy, T.~Onogi, H.~L.~Yu, Phys.~Rev.~{\bf D47}, 2940 
(1993).},
almost all of the top quark mass arises from a new strong ``topcolor''
interaction. To maintain electroweak symmetry between top and bottom quarks
and yet not generate $m_b \simeq m_t$, the topcolor gauge group is
generally taken to be $SU(3)\otimes U(1)$, with the $U(1)$ providing the
difference between top and bottom quarks. Then, in order that topcolor
interactions be natural---i.e., that their energy scale not be far above
$m_t$---and yet not introduce large weak isospin violation, it is necessary
that electroweak symmetry breaking is still due mainly to technicolor
interactions
\ref\tctwohill{C.~T.~Hill, Phys.~Lett.~{\bf 345B}, 483 (1995).}.
In TC2 models, ETC interactions are still needed to generate the light and,
possibly, bottom quark masses, contribute a few~GeV to $m_t$, and give mass
to many technipions. The scale of ETC interactions still must be hundreds
of~TeV to suppress FCNC and, so, the technicolor coupling must still walk.
Two recent papers developing the TC2 scenario are in
Ref.~\ref\tctwoklee{K.~Lane and E.~Eichten, Phys.~Lett.~{\bf B352}, 382
(1995) \semi
K.~Lane, Boston University Preprint BUHEP--96--2, hep-ph/9602221, submitted
to Physical Review~D.}.
Although the phenomenology of TC2 is in its infancy, it is expected to
share general features with multiscale technicolor---many technihadron
states, some carrying ordinary color, some within range of the Tevatron,
and almost all easily produced and detected at the LHC at moderate
luminosities.


\newsec{Signatures for Technicolor and Extended Technicolor}

We assume that the technicolor gauge group is $\sutc$ and that its gauge
coupling walks. A minimal, one-doublet model can have a walking $\atc$ only
if the technifermions belong to a large non-fundamental representation. For
nonminimal models, we generally consider the phenomenology of the lighter
technifermions transforming according to the fundamental~($\Ntc$)
representation; some of these may also be ordinary color triplets. In
almost all respects, walking models are very different from QCD with a few
fundamental $SU(3)$ representations. Thus, arguments based on naive scaling
from QCD and on large-$\Ntc$ certainly are suspect. In TC2, there is no
need for large isospin splitting in the technifermion sector associated
with the top-bottom mass difference. Thus, we can assume negligible
splitting; this simplifies our discussion.

The $\tro \ra W^+W^-$ and $W^\pm Z^0$ signatures of the minimal model were
discussed in Ref.~\ehlq. The principal change due to the large
representation and walking is that scaling the $\tro \ra \tpi\tpi$ coupling
$\atro$ from QCD is questionable. It may be smaller than usually assumed
and lead to a narrower $\tro$. There is also the possibility that, because
of its large mass (naively, 1.5--2~TeV), the $\tro$ has a sizable branching
ratio to four-weak-boson final states. To my knowledge, neither of these
possibilities has been investigated. Enhanced weak-boson pair production in
hadron collisions will be studied at Snowmass by the working group on
Signals for Strong Electroweak Symmetry Breaking.

From now on, we consider only nonminimal models which, we believe, are much
more likely to lead to a satisfactory walking model. They have a rich
phenomenology with many diverse, relatively accessible signals. The masses
of technipions in these models arise from broken ETC and ordinary color
interactions. In walking models we have studied, they lie in the range
100--600~GeV; technirho vector meson masses are expected to lie between 200
and 1000~GeV (see, e.g., Ref.~\multi).


\subsec{Color-Singlet Technipion Production}

Color-singlet technipions, including longitudinal weak bosons $W_L$ and
$Z_L$, are pair-produced via the Drell-Yan process in hadron collisions.
Their $\CO(\alpha^2)$ production rates at the Tevatron and LHC are
unobservably small compared to backgrounds {\it unless} there are fairly
strong color-singlet technirho resonances not far above threshold. To
parameterize the cross sections simply, we consider a model containing two
isotriplets of technipions which mix $W_L^\pm$, $Z_L^0$ with a triplet of
mass-eigenstate technipions $\tpi^{\pm,0}$~\multi,
\ref\tctpi{E.~Eichten and K.~Lane, ``Low-Scale Technicolor at the
Tevatron'', in preparation.}.
We assume that the lighter isotriplet $\tro$ decays into pairs of the state
$\vert\Pi_T\rangle = \sin\chi \ts \vert W_L\rangle + \cos\chi \ts
\vert\tpi\rangle$, leading to the processes
\eqn\singlet{\eqalign{
q \ol q' \ra W^\pm \ra \tropm &\ra \ts\ts W_L^\pm Z_L^0; \quad W_L^\pm
\tpiz, \ts\ts \tpipm Z_L^0; \quad \tpipm \tpiz \cr\cr
q \ol q \ra \gamma, Z^0 \ra \troz &\ra \ts\ts W_L^+ W_L^-; \quad W_L^\pm
\tpimp; \quad \tpip \tpim \ts. \cr}}
The $s$-dependent $\tro$ partial widths are given by (assuming no other
channels, such as colored techipion pairs, are open)
\eqn\singwidth{
\Gamma(\tro \ra \pi_A \pi_B;s) = {2 \atro \CC^2_{AB}\over{3}} \ts
{\ts\ts p_{AB}^3\over {s}} \ts,}
where $p_{AB}$ is the technipion momentum and $\CC^2_{AB} = \sin^4\chi$,
$2\sin^2\chi \cos^2\chi$, $\cos^4\chi$ for $\pi_A \pi_B = W_L W_L$,
$W_L\tpi+\tpi W_L$, $\tpi\tpi$, respectively. The $\tro \ra \tpi\tpi$
coupling $\atro$ obtained by naive scaling from QCD is~\ehlq
\eqn\alpharho{\atro = 2.91 \left({3\over{\Ntc}}\right)\ts.}

Technipion decays are mainly induced by ETC interactions which couple them
to quarks and leptons. These couplings are Higgs-like, and so technipions
are expected to decay into heavy fermion pairs:
\eqn\singdecay{\eqalign{
\tpiz &\ra \cases{b \ol b &if $M_{\tpi} < 2 m_t$,  \cr
t \ol t &if $M_{\tpi} > 2 m_t$; \cr} \cr
\tpip &\ra \cases{c \ol b \ts\ts\ts {\rm or} \ts\ts\ts c \ol s, \ts\ts
\tau^+ \nu_\tau 
&if $M_{\tpi} < m_t + m_b$, \cr
t \ol b &if $M_{\tpi} > m_t + m_b$. \cr} \cr}}
An important caveat to this rule applies to TC2 models. There, only a
few~GeV of the top mass arises from ETC interactions. Then, the $b \ol b$
mode competes with $t \ol t$ for $\tpiz$; $c \ol b$ or $c \ol s$ compete
with $t \ol b$ for $\tpip$. Note that, since the decay $t \ra \tpip b$ is
strongly suppressed in TC2 models, the $\tpip$ can be much lighter than the
top quark.

The $\tro\ra \pi_A \pi_B$ cross sections are well-approximated by
\eqn\singcross{
{d\hat\sigma(q_i \ol q_j \ra \tro^{\pm,0} \ra \pi_A\pi_B) \over{dz}} = 
{\pi \alpha^2 p_{AB}^3 \over{3 \shat^{5/2}}}
\ts {M^4_{\tro} \ts \ts (1-z^2) \over
{(\shat - M_{\tro}^2)^2 + \shat \Gamma_{\tro}^2}} \ts A_{ij}^{\pm,0}(\shat)
\CC^2_{AB} \ts,}
where $\shat$ is the subprocess energy, $z = \cos\theta$ is the
$\pi_A$ production angle, and $\Gamma_{\tro}$ is the $\shat$-dependent
total width of $\tro$. Ignoring Kobayashi-Maskawa mixing angles, the
factors $A_{ij}^{\pm,0} = \delta_{ij} A^{\pm,0}$ are
\eqn\bfactors{\eqalign{
A^\pm &= {1 \over {4 \sin^4\thw}} \biggl({\shat \over {\shat -
M_W^2}}\biggr)^2 \cr
A^0   &= \biggl[Q_i + {2 \cos 2\thw \over {\sin^2 2\thw}} \ts
(T_{3i} - Q_i \sin^2\thw) \biggl({\shat \over {\shat - M_Z^2}}\biggr)
\biggl]^2 \cr 
&\ts + \biggl[Q_i - {2 Q_i \cos 2\thw \sin^2\thw \over{\sin^2
2\thw}} \ts \biggl({\shat \over {\shat - M_Z^2}}\biggr) \biggl]^2 \ts.}}
Here, $Q_i$ and $T_{3i}$ are the electric charge and third component of
weak isospin for~$q_{i L,R}$. Production rates of several picobarns increase
by 5--10 at the LHC; see Table~1.

In the one-family and other models containing colored as well as
color-singlet technifermions, there are singlet and octet technipions
that are electroweak isosinglets commonly denoted $\tpipr$ and $\etat$.
These are singly-produced in gluon fusion. Depending on the technipion's
mass, it is expected to decay to $\ol b b$ (and, possibly, $gg$) or to $\ol
t t$~\ehlq,
\ref\etatrefs{E.~Farhi and L.~Susskind Phys.~Rev.~{\bf D20}
(1979)~3404\semi
S.~Dimopoulos, Nucl.~Phys.~{\bf B168} (1980)~69 \semi
T.~Appelquist and G.~Triantaphyllou, Phys.~Rev.~Lett.~{\bf
69},2750 (1992) \semi
T.~Appelquist and J.~Terning, Phys.~Rev.~{\bf D50}, 2116 (1994)\semi
E.~Eichten and K.~Lane, Phys.~Lett.~{\bf B327}, 129 (1994)\semi
K.~Lane, Phys.~Rev.~{\bf D52}, 1546 (1995).}.
With $\Pi^0 = \tpipr$ or $\etat$, and with constituent technifermions
transforming according to the $\Ntc$~representation of $\sutc$,
the decay rates are
\eqn\Piwidths{\eqalign{
\Gamma(\Pi^0 \ra gg) &= {\CC_\Pi \aqcd^2 \ts \Ntc^2 \ts M_\Pi^3  \over {128
\ts \pi^3 \ts F_T^2}} \ts, \cr\cr
\Gamma(\Pi^0 \ra \ol q q) &= {\gamma_q^2 \ts m_q^2 \ts M_\Pi \ts \beta_q
\over {16 \pi F_T^2}} \ts.\cr}}
Here, $\beta_q = \sqrt{1 - 4m^2_q/M_\Pi^2}$ is the quark velocity. The
$SU(3)$-color factor $\CC_\Pi$ is determined by the triangle-anomaly graph
for $\Pi^0 \ra gg$. In the one-family model, $\CC_\Pi = \tx {4\over{3}}$ for
the singlet $\tpipr$ and $\tx {5\over{3}}$ for the octet $\etat$; values of
$\CO(1)$ are expected in other models. The technipion decay constant $F_T$
is discussed below. The dimensionless factor $\gamma_q$ allows for model
dependence in the technipions' couplings to $\ol q q$. In classical ETC
models, we expect $|\gamma_q| = \CO(1)$. In TC2 models, $|\gamma_q| =
\CO(1)$ for the light quarks and, possibly, the $b$-quark, but $|\gamma_t|
= \CO({\rm few} \ts \gev/m_t) \ll 1$; there will be no $\etat$ enhancement
of $\ol t t$ production in topcolor-assisted technicolor.

The gluon fusion cross section for production and decay of $\Pi^0$ to heavy
$\ol q q$ is isotropic:
\eqn\sigPi{
{d \hat \sigma(gg \ra \Pi^0 \ra \ol q q) \over {d z}} =
{\pi \CN_C \over{32}} \ts {\Gamma(\Pi^0 \ra gg)\ts \Gamma(\Pi^0 \ra \ol q q)
\over {(\shat - M_\Pi^2)^2 + \shat \ts \Gamma^2_{\Pi^0} }} \ts,}
where $\CN_C = 1$ (8) for $\tpipr$ ($\etat$). The decay rates
and cross sections are contolled by the technipion decay constant
$F_T$. In the standard one-family model, $F_T = 123\,\gev$ and the
enhancements in $\ol q q$ production are never large enough to see above
background (unless $\Ntc$ is unreasonably large). In multiscale models and,
we expect, in TC2 models, $F_T$ may be considerably smaller. For example,
in the multiscale model considered in Ref.~\multi, $F_T =30$--$50\,\gev$;
in the TC2 model of Ref.~\tctwoklee, $F_T = 80\,\gev$. Since the total
hadronic cross section,
\eqn\narrowPi{
\sigma(p p^\pm \ra \Pi^0 \ra \ol q q) \simeq {\pi^2 \over {2s}} \ts
{\Gamma(\Pi^0 \ra gg) \ts \Gamma(\Pi^0 \ra \ol q q) \over {M_\Pi \ts
\Gamma_{\Pi^0}}}
\ts \int d \eta_B \ts f_g^p\biggl({M_\Pi\over{\sqrt{s}}}
e^{\eta_B}\biggr)
\ts f_g^p\biggl({M_\Pi\over{\sqrt{s}}} e^{-\eta_B}\biggr) \ts,}
scales as $1/F_T^2$, small decay constants may lead to observable
enhancements in $\ol t t$ production in standard multiscale technicolor and
in $\ol b b$ production in TC2. Sample rates are given in Table~1.

In models containing colored technifermions, color-singlet technipions are
also pair-produced in the isospin $I=0$ channel via gluon fusion. This
process involves intermediate states of color-triplet and octet
technipions. Again, the subprocess cross section is isotropic; it is given by
\ref\tpitev{K.~Lane, Phys.~Lett.~{\bf B357}, 624 (1995)\semi also see
T.~Lee, Talk presented at International Symposium on Particle Theory and
Phenomenology, Ames, IA, May 22-24, 1995, FERMILAB-CONF-96-019-T,
hep-ph/9601304, (1996).}.
\eqn\dsggpp{\eqalign{
& {d \hat\sigma(gg \ra \tpip \tpim) \over{dz}} =
2{d \hat\sigma(gg \ra \tpiz \tpiz) \over{dz}} \cr
& \qquad = {\aqcd^2 \beta \over {2^{15} \pi^3 F_T^4 \shat}}
\ts \biggl\vert T(R) \ts \left[C_R \ts \left(\shat  - \twothirds(2M_R^2
+ M_{\pi_T}^2)\right) + D_R \right] \ts \left(1 +
2\CI(M_R^2,\shat)\right)\biggl\vert^2 \ts. \cr }}
Here, $\beta = 2p/\rshat$ is the technipion velocity.
The sum is over $SU(3)$ representations $R =3,8$ of the $\tpi$ and $T(R)$
is the trace of the square of their $\suc$-generator matrices: $T(R) =
\half$ for triplets (dimension $d(R) = 3$), 3 for octets ($d(R) = 8$). The
factors $C_R$ and $D_R$ are listed in Table~2 for the one--family model and
a multiscale model. The integral~$\CI$ is
\eqn\zint{\eqalign{
\CI(M^2,s) &\equiv  \int_0^1 dx \ts dy \ts {M^2 \over {xys - M^2 +
i\epsilon}} \ts \theta(1-x-y) \cr
&=\cases{-M^2 /2s \left[ \pi - 2 \arctan \sqrt{4 M^2/s -1}
\right]^2 & for $s <  4M^2$ \cr
M^2/2s \left[ \ln \left({1 + \sqrt{1 - 4 M^2/s} \over
{1 - \sqrt{1 - 4 M^2/s}}}\right) - i\pi\right]^2  &for $s > 4M^2$ \ts.}
\cr}}
The rates at the Tevatron are at most comparable to those enhanced by
technirhos; they are considerably greater at the the LHC because the fusing
gluons are at low~$x$ (see Table~1). An interesting feature of this cross
section is that the $\tpi\tpi$ invariant mass distribution peaks near the
color-triplet and octet technipion thresholds, which can be well above $2
M_{\tpi}$. It is possible that mixed modes such as $W_L^\pm \tpimp$ and
$Z_L \tpiz$ are also produced by gluon fusion, with the rates involving
mixing angles such as $\chi$ in Eq.~\singcross.


\subsec{Color-Octet Technirho  Production and Decay to Jets and Technipions}

Models with an electroweak doublet of color-triplet techniquarks $(U,D)$
have an octet of $I=0$ technirhos, $\troct$, with the same quantum numbers
as the gluon. The $\troct$ are produced strongly in $\ol q q$ and $gg$
collisions. Assuming, for simplicity, one doublet $(N,E)$ of color-singlet
technileptons (as in the one-family model), there are the 63 technipions
listed in Section~2. The color-singlet and octet technipions decay as in
Eq.~\singdecay\ above. The leptoquark decay modes are expected to be
\eqn\leptoquark{\eqalign{
\tpiun &\ra \cases{c \ol \nu_\tau &if $M_{\tpi} <  m_t$,  \cr
t \ol \nu_\tau &if $M_{\tpi} > m_t$; \cr} \cr
\tpiue &\ra \cases{c \tau^+ &if $M_{\tpi} <  m_t$,  \cr
t \tau^+ &if $M_{\tpi} > m_t$; \cr} \cr
\tpidn &\ra b \ol \nu_\tau \ts; \cr
\tpide &\ra b \tau^+ \ts.  \cr}}
The caveat regarding technipion decays to top quarks in TC2 models still
applies.

There are two possibilities for $\troct$ decays~\multi. If walking
technicolor enhancements of the technipion masses close off the $\tpi\tpi$
channels, then $\troct \ra \ol q q,\ts gg \ra \jets$. The color-averaged
$\CO(\aqcd^2)$ cross sections are given by
\eqn\dsjets{\eqalign{
& {d\hat \sigma(\ol q_i q_i \ra \ol q_i q_i) \over {d z}} = 
{2 \pi \aqcd^2 \over {9 \shat}} \left\{ \ts \bigl| \CDgg(\shat) \bigr|^2 \ts
\left({\uhat^2 + \that^2 \over {\shat^2}} \right)
- \twothirds \ts {\rm Re}\ts \CDgg(\shat) \ts \left({\uhat^2 \over
{\shat\that}}\right)
+ {\shat^2 + \uhat^2 \over {\that^2}} \right\}
\ts; \cr\cr
& {d\hat \sigma(\ol q_i q_i \ra \ol q_j q_j)
\over {dz}} = {2 \pi \aqcd^2 \over {9 \shat}} \ts \bigl| \CDgg(\shat)
\bigr|^2 \ts \left({\uhat^2 + \that^2 \over {\shat^2}} \right)
\ts; \cr\cr
& {d\hat \sigma(\ol q_i q_i \ra gg) \over {d z}} =
{64 \over {9 }} {d\hat \sigma(gg \ra q_i \ol q_i) \over {d z}} 
= {4 \pi \aqcd^2 \over {3 \shat}}
\left\{ \ts \bigl| \CDgg(\shat) - 1 \bigr|^2 \ts
{2\uhat\that \over{\shat^2}}
+ \tx{{4 \over {9}}} \biggl({\uhat \over {\that}} + {\that \over {\uhat}}
\biggr) 
- {\uhat^2 + \that^2 \over {\shat^2}} \right\}
\ts; \cr\cr
& {d\hat \sigma(gg \ra gg) \over {d z}} = {9 \pi \aqcd^2 \over {4 \shat}}
\biggl\{ \ts 3 - {\uhat\that\over{\shat^2}} - {\that\shat\over{\uhat^2}} 
- {\shat\uhat\over{\that^2}} \cr
& \qquad\qquad\qquad + \tx{{1 \over {4}}} \bigl| \CDgg(\shat) - 1 \bigr|^2
\ts\left({\uhat - \that \over{\shat}}\right)^2
- \tx{{1 \over {4}}} {\rm Re}(\CDgg(\shat) -1) \ts 
{\left(\uhat - \that\right)^2\over {\uhat\that}}
\biggr\}
\ts; \cr\cr
& {d\hat \sigma(q_i q_j \ra q_i q_j) \over {d z}} =
{d\hat \sigma(\ol q_i \ol q_j \ra \ol q_i \ol q_j) \over {d z}}
= {d\hat \sigma(q_i \ol q_j \ra q_i \ol q_j) \over {d z}}
= {2 \pi \aqcd^2 \over {9 \shat}} \ts
\left({\shat^2 + \uhat^2 \over {\that^2}}\right)
\ts; \cr\cr
& {d\hat \sigma(q_i q_i \ra q_i q_i) \over {d z}} =
{d\hat \sigma(\ol q_i \ol q_i \ra \ol q_i \ol q_i) \over {d z}}
= {2 \pi \aqcd^2 \over {9 \shat}} \ts \left\{
{\shat^2 + \uhat^2 \over{\that^2}} + {\that^2 + \uhat^2 \over{\shat^2}}
- \twothirds {\shat^2 \over {\uhat\that}} \right\} \ts;\cr \cr
& {d\hat \sigma(g q_i \ra g q_i) \over {d z}} =
{d\hat \sigma(g \ol q_i \ra g \ol q_i) \over {d z}}
= {\pi \aqcd^2 \over {2 \shat}} \ts (\shat^2 + \uhat^2) \ts
\left( \ts {1 \over {\that^2}} - {4 \over {9 \shat\uhat}} \right)
\ts . \cr}}
Here, $z = \cos\theta$, $\that = -\half \shat(1 - z)$, $\uhat = -\half
\shat(1 + z)$ and it is understood that $q_i \ne q_j = u,d,c,s,b$
contribute to dijet events. Only the $s$-channel gluon propagator was
modified to include the $\troct$ resonance. Here and below, we use the
dimensionless propagator factors $\CDgg$ and $\CDgrho$
\eqn\cdg{\eqalign{
\CDgg(s) &= {s - M_{\troct}^2 + i \ecm \ts \Gamma_{\troct}(s) \over
{s(1- 2\aqcd(s)/\atro) - M_{\troct}^2 + i \ecm \ts \Gamma_{\troct}(s)}}
\ts, \cr\cr
\CDgrho(s) &= {s \over
{s(1- 2\aqcd(s)/\atro) - M_{\troct}^2 + i \ecm \ts
\Gamma_{\troct}(s)}} \ts. \cr}}
The $s$-dependent $\troct$ width in this case is the sum of (allowing for
multijet $\ol t t$ final states, assumed light compared to $\sqrt{s}$)
\eqn\troctwidth{\eqalign{
&\sum_{i=1}^6 \Gamma(\troct \ra \ol q_i q_i) = {6 \over {3}} {\aqcd^2(s)
\over {\atro}} \ecm \ts , \cr
&\Gamma(\troct \ra gg) = {\aqcd^2(s) \over {\atro}} \ecm \ts . \cr }}
A search for the dijet signal of $\troct$ has been carried out by the CDF
Collaboration; see Ref.~\ref\cdfdijet{F.~Abe, et al., The CDF
Collaboration, Phys.~Rev.~Lett.~{\bf 74}, 3538 (1995).}
for a detailed discussion of expected signal and background rates.
Rough signal-to-background estimates are given in Table~1. They are sizable
at the Tevatron and LHC, but are sensitive to jet energy resolutions. 

Colored technipions are pair-produced in hadron collisions through
quark-antiquark annihilation and gluon fusion. If the $\troct \ra
\tpi\tpi$ decay channels are open, this production is resonantly
enhanced. The subprocess cross sections, averaged over initial colors
and summed over the colors $B$, $C$ of technipions, are given by
\eqn\qbqpipi{\sum_{B,C} {d\hat \sigma(\ol q_i q_i \ra \pi_B \pi_C) \over
{dz}} = {\pi \aqcd^2(\shat) \beta^3 \over {9 \shat}} \ts \CS_{\pi} T(R) \ts
\bigl(1 - z^2 \bigr) \ts \bigl| \CDgg + \CDgrho \bigr|^2 \ts,}
\eqn\ggpipi{\eqalign{
&\sum_{B,C}{d\hat \sigma(gg \ra \pi_B \pi_C)\over {dz}} =
 {\pi \aqcd^2(\shat) \beta \over {\shat}} \ts
\CS_{\pi} T(R) \biggl\{{3 \over {32}} \ts \beta^2 \ts z^2 \ts
\biggl[\bigl| \CDgg + \CDgrho \bigr|^2 \cr
&\qquad\qquad -{2 \beta^2 \ts (1-z^2) \over {1-\beta^2 z^2}}
\ts {\rm Re}\ts \left(\CDgg + \CDgrho \right)
+ 2 \biggl({\beta^2 \ts (1-z^2)
\over {1 - \beta^2 z^2}} \biggr)^2 \biggr] \cr
&\qquad \qquad + \left({T(R) \over {d(R)}} - {3 \over {32}} \right)
\biggl[ {(1 - \beta^2)^2 + \beta^4 \ts (1-z^2)^2
\over {(1 - \beta^2 z^2)^2}} \biggr] \biggr\}
\ts,\cr}}
where $\beta$ is the technipion velocity and $z = \cos\theta$.
The symmetry factor $\CS_{\pi} = 1$ for each channel of $\tpilq \tpiql$ and
for $\octpip \octpim$; $\CS_{\pi} = \half$ for the identical-particle final
states, $\octpiz \octpiz$ and $\etat\etat$. The $SU(3)$ group factors
$T(R)$ and $d(R)$ for $R=3,8$ were defined above at Eq.~\dsggpp. The
technirho width is now the sum of the $\ol q q$ and $gg$ partial widths and
\eqn\octwidth{
\sum_{B,C} \Gamma(\tro \ra \pi_B \pi_C;s) = {\atro \CS_\pi T(R) \over {3}}
{\ts\ts\ts p^3 \over{s}} \ts.}
As indicated in Table~1, pair-production rates for colored technipions with
masses of a few hundred~GeV are several picobarns at the Tevatron, rising to
a few nanobarns at the LHC.


\subsec{Signatures of Topcolor-Assisted Technicolor}

The development of topcolor-assisted technicolor is still at an early stage
and, so, its phenomenology is not fully formed. Nevertheless, there are
three TC2 signatures that are likely to be present in any surviving
model~\topcondref--\tctwoklee,
\ref\hp{C.~T.~Hill and S.~Parke, Phys.~Rev.~{\bf D49}, 4454 (1994)\semi
Also see K.~Lane, Phys.~Rev.~{\bf D52}, 1546 (1995)\semi I thank
D.~Kominis for corrections to a numerical errors in both papers.}:

\item{$\circ$} The isotriplet of color-singlet ``top-pions'' $\toppi$ arising
from spontaneous breakdown of the top quark's $SU(2)\otimes U(1)$ chiral
symmetry;

\item{$\circ$} The color-octet of vector bosons $V_8$, called ``colorons'',
associated with breakdown of the top quark's strong $SU(3)$ interaction to
ordinary color;

\item{$\circ$} The $Z'$ vector boson associated with breakdown of the top
quark's strong $U(1)$ interaction to ordinary weak hypercharge.

\medskip

The three top-pions are nearly degenerate. They couple to the top quark
with strength $m_t/F_t$, where $m_t$ is the part of the top-quark mass
induced by topcolor---within a few GeV of its total mass---and $F_t \simeq
70\,\gev$~\tctwohill\ is the $\toppi$ decay constant.\foot{As far as I
know, the rest of the discussion in this and the next paragraph has not
appeared in print before. It certainly deserves more thought than has gone
into it here. One possible starting place is the paper by Hill, Kennedy,
Onogi and Yu in Ref.~\topcref.} If
the top-pion is lighter than the top quark, then
\eqn\tpibrate{\Gamma(t \ra \toppip b) =
{(m_t^2 - M_{\pi_t}^2)^2 \over {32 \pi m_t F_t^2}} \ts.}
It is known that $B(t \ra W^+ b) = 0.87\pm^{+0.13}_{-0.30}$ (stat.)
$^{+0.13}_{-0.11}$ (syst.)
\ref\twbrate{J.~Incandela, Proceedings of the 10th Topical Workshop on
Proton-Antiproton Collider Physics, Fermilab, edited R.~Raja and J.~Yoh,
p.~256 (1995).}.
At the $1\sigma$ level, then, $M_{\pi_t} \simge 150\,\gev$. At the
$2\sigma$ level, the lower bound is $100\,\gev$, but such a small branching
ratio for $t \ra W^+ b$ would require $\sigma(p\ol p \ra t \ol t)$ at the
Tevatron about~4 times the standard QCD value of
$4.75^{+0.63}_{-0.68}\,\pb$
\ref\topsig{S.~Catani, M.~Mangano, P.~Nason and L.~Trentadue,
CERN-TH/96-21, hep-ph/9602208 (1996).}.
The $t \ra \toppip b$ decay mode can be sought in high-luminosity runs
at the Tevatron and with moderate luminosity at the LHC. If $M_{\pi_t} <
m_t$, then $\toppip \ra c \ol b$ through $t$--$c$ mixing. It is also
possible, though unlikely, that $\tpip \ra t \ol s$ through $b$--$s$
mixing.

If $M_{\pi_t} > m_t$, then $\toppip \ra t \ol b$ and $\toppiz \ra \ol t t$
or $\ol c c$, depending on whether the top-pion is heavier or lighter than
$2m_t$. The main hope for discovering top-pions heavier than the top quark
seems to rest on the isotriplet of top-rho vector mesons, $\rho_t^{\pm,0}$.
It is hard to estimate $M_{\rho_t}$; it may lie near $2m_t$ or closer to
$\Lambda_t = \CO(1\,\tev)$. They are produced in hadron collisions just as
the corresponding color-singlet technirhos (Eq.~\singlet). The conventional
expectation is that they decay as $\rho_t^{\pm,0} \ra \toppipm \toppiz$,
$\toppip \toppim$. Then, the top-pion production rates may be estimated
from Eqs.~\singwidth\ and \singcross\ with $\atro = 2.91$ and $\CC_{AB} =
1$. The rates are not large, but the distinctive decays of top-pions help
suppress standard model backgrounds.

Life may not be so simple, however. The $\rho_t$ are not completely
analogous to the $\rho$-mesons of QCD and technicolor because topcolor is
broken near $\Lambda_t$. Thus, for distance scales between $\Lambda^{-1}_t$
and $1\,\gev^{-1}$, top and bottom quarks do not experience a growing
confining force. Instead of $\rho_t \ra \toppi\toppi$, it is also possible
that $\rho_t^{\pm,0}$ fall apart into their constituents $t \ol b$, $b \ol
t$ and $t \ol t$. The $\rho_t$ resonance may be visible as a significant
increase in $t \ol b$ production, but it won't be in $t \ol t$.\foot{I
thank John Terning for inspiring this discussion of $\rho_t$ decays.}

The $V_8$ colorons of broken $SU(3)$ topcolor are readily produced in
hadron collisions. They are expected to have a mass between 1/2--1~TeV.
Colorons couple with strength $-g_S \cot\xi$ to quarks of the two light
generations and with strength $g_S \tan\xi$ to top and bottom quarks, where
$\tan\xi \gg 1$~\hp. Their decay rate is
\eqn\widveight{
\Gamma_{V_8} = {\aqcd \Mv \over {6}}\ts
\biggl\{4\cot^2\xi + \tan^2\xi \left(1 + \beta_t
(1-m_t^2/\Mv^2)\right)\biggr\} \ts.}
where $\beta_t = \sqrt{1 - 4 m_t^2/\Mv^2}$. Colorons may then appear as
resonances in $b \ol b$ and $t \ol t$ production. For example, the
$\CO(\aqcd)$ cross section for $\ol q q \ra \ol t t$ becomes
\eqn\qqbttb{
{d \hat \sigma(\ol q q\ra \ol t t) \over {d z}} =
{\pi \aqcd^2 \beta_t \over {9 \shat}} \ts \bigl(2 - \beta_t^2 +
\beta_t^2 z^2\bigr) \ts \biggl|1 - \ts{\shat \over
{\shat - \Mv^2 + i \sqrt{\shat} \ts \Gamma_{V_8}}}\biggr|^2 \ts.}
For completeness, the $gg \ra \ol t t$ rate is
\eqn\ggttb{\eqalign{
{d \hat \sigma(gg \ra \ol t t) \over {d z}} =
{\pi \aqcd^2 \beta_t \over {6 \shat}} \ts
&\biggl\{{1 + \beta_t^2 z^2 \over {1 - \beta_t^2 z^2}} -
{(1-\beta_t^2)^2 \ts (1 + \beta_t^2 z^2) \over{(1-\beta_t^2 z^2)^2}}
- \tx{{9\over{16}}} (1 + \beta_t^2 z^2) \cr
& + {1-\beta_t^2 \over{1-\beta_t^2 z^2}} \ts (1 -
\tx{{1\over{8}}} \beta_t^2 + \tx{{9\over{8}}} \beta_t^2 z^2)
\biggr\} \ts. \cr}}
A description of the search for colorons and other particles decaying
to $\ol b b$ and $\ol t t$ and preliminary limits on their
masses are given in Ref.~\ref\cdfcoloron{R.~M.~Harris, Proceedings of the
10th Topical Workshop on Proton-Antiproton Collider Physics, Fermilab,
edited R.~Raja and J.~Yoh, p.~72(1995).}.

Colorons have little effect on the standard dijet production rate. The
situation is very different for the $Z'$ boson of the broken strong $U(1)$
interaction.\foot{This interaction differentiates between top and bottom
quarks, helping the former develop a large mass while keeping the latter
light.} In Ref.~\tctwoklee\ a scenario for topcolor was developed in which
it is necessary that the $Z'$ couples strongly to the fermions of the first
two generations as well as those of the third. The $Z'$ probably is heavier
than the colorons, roughly $\Mzp =1$--$2\,\tev$. Thus, at subprocess
energies well below $\Mzp$, the interaction of $Z'$ with all quarks is
described by a contact interaction, just what is expected for quarks with
substructure at the scale $\Lambda \sim 1$--$2\,\tev$. This leads to an
excess of jets at high $E_T$ and invariant mass
\ref\elp{E.~J.~Eichten, K.~Lane and M.~E.~Peskin, Phys.~Rev.~Lett.~{\bf
50}, 811 (1983)},\ehlq.
An excess in the jet-$E_T$ spectrum consistent with $\Lambda = 1600\,\gev$
has been reported by the CDF Collaboration
\ref\cdfjets{F.~Abe, et al., The CDF Collaboration, FERMILAB-PUB-96/020-E,
hep-ex/9601008, submitted to Physical Review Letters (1996).}.
It remains to be seen whether it is due to topcolor or any other new
physics. As with quark substructure, the angular and rapidity distributions
of the high-$E_T$ jets induced by $Z'$ will be more central than predicted
by QCD. The $Z'$ will also produce an excess of high invariant mass
$\ell^+\ell^-$. It will be interesting to compare limits on contact
interactions in the Drell-Yan process with those obtained from jet
production.

The topcolor $Z'$ will be produced directly in $\ol q q$ annhilation in LHC
experiments. Because the $Z'$ is strongly coupled to so many fermions,
including technifermions in the LHC's energy range, it is likely to be very
broad. The development of TC2 models is at such an early stage that the
$Z'$ couplings, its width and branching fractions, cannot be predicted with
confidence. These studies are underway and we can expect progress on these
questions in the coming year.


\newsec{Signatures for Quark and Lepton Substructure}

The presence of three generations of quarks and leptons, apparently
identical except for mass, strongly suggests that they are composed of
still more fundamental fermions, often called ``preons''. It is clear that,
if preons exist, their strong interaction energy scale $\Lambda$ must be
much greater than the quark and lepton masses. Long ago, 't~Hooft figured
out how interactions at high energy could produce essentially massless
composite fermions: the answer lies in unbroken chiral symmetries of the
preons {\it and} confinement by their strong ``precolor'' interactions
\ref\thooft{G.~'t~Hooft, in {\it Recent Developments in
Gauge Theories}, edited by G.~'t~Hooft, et al. (Plenum, New York, 1980).}.
There followed a great deal of theoretical effort to construct a realistic
model of composite quarks and leptons (see, e.g.,
Ref.~\ref\comp{S.~Dimopoulos, S.~Raby and L.~Susskind, Nucl.~Phys.~{\bf
B173}, 208 (1980)\semi
M.~E.~Peskin, Proceedings of the 1981 Symposium on Lepton and Photon
Interactions at High Energy, edited by W.~Pfiel, p.~880 (Bonn, 1981) \semi
I.~Bars, Proceedings of the Rencontres de Moriond, {\it Quarks, Leptons and
Supersymmetry}, edited by Tranh Than Van, p.~541 (1982).}) which,
while leading to valuable insights on chiral gauge theories, fell far short
of its main goal.

In the midst of this activity, it was pointed out that the existence of
quark and lepton substructure will be signalled at energies well below
$\Lambda$ by the appearance of four-fermion ``contact'' interactions which
differ from those arising in the standard model~\elp. These interactions
are induced by the exchange of preon bound states and precolor-gluons. They
must be $SU(3) \otimes SU(2) \otimes U(1)$ invariant because they are
generated by forces operating at or above the electroweak scale. These
contact interactions are suppressed by $1/\Lambda^2$, but the coupling
parameter of the exchanges---analogous to the pion-nucleon and rho-pion
couplings---is not small. Thus, the strength of these interactions is
conventionally taken to be $\pm 4\pi/\Lambda^2$. Compared to the standard
model, contact interaction amplitudes are then of relative order
$\shat/\aqcd \Lambda^2$ or $\shat/\alpha_{EW} \Lambda^2$. The appearance of
$1/\alpha$ and the growth with $\shat$ make contact-interaction effects the
lowest-energy signal of quark and lepton substructure. They are sought in
jet production at hadron and lepton colliders, Drell-Yan production of high
invariant mass lepton pairs, Bhabha scattering, $e^+e^- \ra \mu^+\mu^-$ and
$\tau^+\tau^-$
\ref\pdg{For current collider limits on substructure, see Ref.~\cdfjets\
and the Review of Particle Properties, Particle Data Group, {\it
Phys.~Rev.}~{\bf D50}, 1173, (1992).},
atomic parity violation
\ref\rosner{J.~Rosner, Phys.~Rev.~{\bf D53}, 2724 (1996), and references
therein.},
and polarized M{\o}ller scattering
\ref\kumar{K.~Kumar, E.~Hughes, R.~Holmes and P.~Souder, ``Precision Low
Energy Weak Neutral Current Experiments'', Princeton University
(October 30, 1995), to appear in Modern Physics Letters~A.}.
Here, we concentrate on jet production and the Drell-Yan process at hadron
colliders.

The contact interaction most used so far to parameterize limits on the
substructure scale $\Lambda$ is the product of left-handed electroweak
isoscalar quark and lepton currents. Collider experiments can probe values
of $\Lambda$ in the 2--5~TeV range (Tevatron) to the 15--20~TeV range (LHC;
see Refs.~\ehlq\ and \ref\gemtdr{GEM Technical Design Report, Chapter~2.6,
GEM~TN-93-262, SSCL-SR-1219; submitted to the Superconducting Super
Collider Laboratory, (April 30, 1993)\semi K.~Lane, F.~Paige, T.~Skwarnicki
and J.~Womersley, {\it Simulations of Supercollider Physics}, Boston
University preprint BUHEP-94-31, Brookhaven preprint BNL-61138,
hep-ph/9412280 (1994), to appear in Physics Reports.}). If $\Lambda$ is to
be this low, the contact interaction must be flavor-symmetric, at least for
quarks in the first two generations, to avoid large $\Delta S = 2$ and,
possibly, $\Delta B_d = 2$ neutral current interactions. We write it, the
only Lagrangian we exhibit, as
\eqn\Lcomp{
\CL^0_{LL} = {4 \pi \eta \over {2 \Lambda^2}}
\sum_{i,j=1}^3 \ts \left( \sum_{a=1}^3 \ol q_{aiL} \gamma^\mu q_{aiL} +
\CF_\ell \ts \ol \ell_{iL} \gamma^\mu \ell_{iL} \right)
\ts \left( \sum_{b=1}^3 \ol q_{bjL} \gamma_\mu q_{bjL} +
\CF_\ell \ts \ol \ell_{jL} \gamma_\mu \ell_{jL} \right) \ts.}
Here, $\eta=\pm 1$; $a,b = 1,2,3$ labels color; $i,j = 1,2,3$ labels
the generations, and the quark and lepton fields are isodoublets,
$q_{ai} = (u_{ai}, d_{ai})$ and $\ell_i = (\nu_i, e_i)$ (with right-handed
neutrinos generally omitted). The real factor $\CF_\ell$ is inserted to
allow for different quark and lepton couplings, but it is expected to be
$\CO(1)$. The factor of $\half$ in the overall strength of the interaction
avoids double-counting interactions and amplitudes.

The color-averaged jet subprocess cross sections, modified for the
interaction $\CL^0_{LL}$, are given in leading order in $\aqcd$ by (these
formulas correct errors in Ref.~\ehlq)
\eqn\sigcompjet{\eqalign{
& {d \hat\sigma(q_i q_i \ra q_i q_i) \over {dz}} =
{d \hat\sigma(\ol q_i \ol q_i \ra \ol q_i \ol q_i) \over {dz}} \cr
& = {\pi \over {2 \shat}}
\biggl\{ {4 \over {9}} \aqcd^2 \left[{\uhat^2 + \shat^2 \over {\that^2}} +
{\that^2 + \shat^2 \over {\uhat^2}} -
{2 \over {3}} {\shat^2 \over {\that \uhat}} \right]
+ {8 \over {9}} \aqcd {\eta \over {\Lambda^2}} \left[ {\shat^2 \over
{\that}} + {\shat^2 \over {\uhat}} \right] + {8 \over {3}} {\shat^2 \over
{\Lambda^4}} \biggr\} \ts; \cr\cr
& {d \hat\sigma(q_i \ol q_i \ra q_i \ol q_i) \over {dz}}\cr
& = {\pi \over {2 \shat}}
\biggl\{ {4 \over {9}} \aqcd^2 \left[{\uhat^2 + \shat^2 \over {\that^2}} +
{\uhat^2 + \that^2 \over {\shat^2}} -
{2 \over {3}} {\uhat^2 \over {\shat \that}} \right]
+ {8 \over {9}} \aqcd {\eta \over {\Lambda^2}} \left[ {\uhat^2 \over
{\that}} + {\uhat^2 \over {\shat}} \right] + {8 \over {3}} {\uhat^2 \over
{\Lambda^4}} \biggr\} \ts; \cr\cr
& {d \hat\sigma(q_i \ol q_i \ra q_j \ol q_j) \over {dz}} = 
{\pi \over {2 \shat}}
\biggl\{ {4 \over {9}} \aqcd^2 \left[{\uhat^2 + \that^2 \over {\shat^2}}
\right] + {\uhat^2 \over {\Lambda^4}} \biggr\} \ts; \cr\cr
& {d \hat\sigma(q_i \ol q_j \ra q_i \ol q_j) \over {dz}} =
{\pi \over {2 \shat}}
\biggl\{ {4 \over {9}} \aqcd^2 \left[{\uhat^2 + \shat^2 \over {\that^2}}
\right] + {\uhat^2 \over {\Lambda^4}} \biggr\} \ts; \cr\cr
& {d \hat\sigma(q_i q_j \ra q_i q_j) \over {dz}} =
{d \hat\sigma(\ol q_i \ol q_j \ra \ol q_i \ol q_j) \over {dz}} =
{\pi \over {2 \shat}}
\biggl\{ {4 \over {9}} \aqcd^2 \left[{\uhat^2 + \shat^2 \over {\that^2}}
\right] + {\shat^2 \over {\Lambda^4}} \biggr\}\ts .\cr}}
For this LL-isoscalar interaction, the interference term ($\eta/\Lambda^2$)
in the hadron cross section is small and the sign of $\eta$ is not very
important. Interference terms may be non-negligible in contact interactions
with different chiral, flavor, and color structures. In all cases, the main
effect of substructure is to increase the rate of centrally-produced jets.
Seeing this in the jet angular distribution is important for confirming the
presence of contact interactions.

The modified cross sections for the Drell-Yan process $\ol q_i q_i \ra
\ell_j^+ \ell_j^-$ is
\eqn\sigcomplept{
{d\hat\sigma(\ol q_i q_i \ra \ell_j^+ \ell_j^-) \over {dz}} = {\pi
\alpha^2 \over {6 \shat}} \ts \left[ \CA_i(\shat) \left({\uhat \over
{\shat}} \right)^2 +
\CB_i(\shat) \left({\that \over {\shat}} \right)^2 \right] \ts ,}
where
\eqn\cacb{\eqalign{
\CA_i(\shat) = &\biggl[ Q_i + {4 \over {\sin^2 2 \thw}}
\left(T_{3i} - Q_i \sin^2 \thw \right) \left(\half - \sin^2 \thw
\right)
\left({\shat \over {\shat - M_Z^2}}\right) - {\CF_\ell \eta \shat \over
{\alpha \Lambda^2}} \biggr]^2 \cr
+&\biggl[ Q_i + Q_i \tan^2 \thw \left({\shat \over {\shat - M_Z^2}}\right)
\biggr]^2 \ts ; \cr
\CB_i(\shat) = &\biggl[ Q_i - {1 \over{\cos^2 \thw}} 
\left(T_{3i} - Q_i \sin^2 \thw \right) \left({\shat \over {\shat - M_Z^2}}
\right) \biggr]^2 \cr
+&\biggl[ Q_i - {1 \over {\cos^2 \thw}} Q_i  \left(\half - \sin^2
\thw \right) \left({\shat \over {\shat - M_Z^2}}\right) \biggr]^2 \ts.
\cr}}

The angular distribution of the $\ell^-$ relative to the incoming quark is
an important probe of the contact interaction's chiral structure. Measuring
this distribution is easy in a $\ol p p$ collider such as the Tevatron
since the hard quark almost always follows the proton direction. If the
scale $\Lambda$ is high so that parton collisions revealing the contact
interaction are hard, the quark direction can also be determined with
reasonable confidence in a $pp$ collider. At the LHC, the quark in a $\ol q
q$ collision with $\sqrt{\shat/s} \simge 1/20$ is harder than the
antiquark, and its direction is given by the boost rapidity of the dilepton
system, at least 75\% of the time. The charges of $\CO(1\,\tev)$ muons can
be well-measured even at very high luminosity in the detectors being
designed for the LHC. These two ingredients are needed to insure a good
determination of the angular distribution~\gemtdr.

It is important to study the effects of contact interactions with chiral,
flavor and color structures different from the one in Eq.~\Lcomp. Such
interactions can give rise to larger (or smaller) cross sections for the
same $\Lambda$ because they have more terms or because they interfere more
efficiently with the standard model. Thus, it will be possible to probe
even higher values of $\Lambda$ for other structures. Other forms can also
give rise to $\ell^\pm\nu$ final states. Searching for contact interactions
in these modes is more challenging than in $\ell^+\ell^-$, but it is very
useful for untangling flavor and chiral structures~\gemtdr. Events are
selected which contain a single high-$p_T$ charged lepton, large missing
$E_T$ and little jet activity. Even though the parton c.m.~frame cannot be
found in this case, it is still possible to obtain information on the
chiral nature of the contact interaction by comparing the rapidity
distributions, $\vert \eta_{\ell^+} \vert$ and $\vert \eta_{\ell^-} \vert$
of the high-$p_T$ leptons. For example, if the angular distribution in the
process $d \ol u \ra \ell^- \ol \nu$ between the incoming $d$-quark and the
outgoing $\ell^-$ is $(1 + \cos\theta)^2$, then $\vert\eta_{\ell^-}\vert$
is pushed to larger values because the $d$-quark is harder than the $\ol
u$-quark and the $\ell^-$ tends to be produced forward. Correspondingly,
in $u \ol d \ra \nu \ell^+$, the $\vert\eta_{\ell^+}\vert$ distribution
would be squeezed to smaller values.


\newsec{Conclusions and Acknowledgements}

Many theorists are convinced that low-energy supersymmetry is intimately
connected with electroweak symmetry breaking and that its discovery is just
around the corner
\ref\susylett{J.~Bagger, et al., Letter to the Director of Fermilab and
the Co-Spokespersons of the CDF and D\O\ Collaborations, (1994).}.
One often hears that searches for other TeV-scale physics are a waste of
time. Experimentalists know better. The vast body of {\it experimental}
evidence favors no particular extension of the standard model. Therefore,
all plausible approaches must be considered. Detectors must have the
capability---and experimenters must be prepared---to discover whatever
physics is responsible for electroweak and flavor symmetry breaking. To
this end, we have summarized the principal signatures for technicolor,
extended technicolor and quark-lepton substructure. Table~1 lists sample
masses for new particles and their production rates at the Tevatron and
LHC. We hope this summary helps in-depth studies of strong TeV-scale
dynamics get underway at Snowmass.

\medskip

This Snowmass catalog was suggested by Ian Hinchliffe. I am especially
grateful to John Womersley for encouragement, advice and a thoughtful
reading of the manuscript. I owe a great deal to Dave Cutts, Robert Harris,
Kaori Maeshima and Wyatt Merritt for their guide to the physicists in D\O\
and CDF who are carrying out searches for the signatures of strong
electroweak and flavor dynamics. These people are a valuable resource whose
wisdom and experiences will be indispensible at Snowmass and I thank them
for their willingness to help. I am also indebted to those members of CDF
and D\O\ who discussed their work with me and otherwise made my task
possible: Tom Baumann, Paul Grannis, John Huth, Hugh Montgomery and Jorge
Troconiz. Finally, I thank Dimitris Kominis for discussions on
topcolor-assisted technicolor and for catching several errors.

\vfil\eject

\listrefs

\centerline{\vbox{\offinterlineskip
\hrule\hrule
\halign{&\vrule#&
  \strut\quad#\hfil\quad\cr\cr
height4pt&\omit&&\omit&&\omit&&\omit&\cr\cr
&\hfill Process \hfill&&\hfill
Sample Mass (GeV)  \hfill&&\hfill $\sigma_{\rm TeV}$(pb) \hfill&&\hfill
$\sigma_{\rm LHC}$(pb) \hfill &\cr\cr
height4pt&\omit&&\omit&&\omit&&\omit&\cr\cr
\noalign{\hrule\hrule}
height4pt&\omit&&\omit&&\omit&&\omit&&\omit&&\omit&\cr\cr
&$\tro\ra W_L\tpi${$\ts ^{1}$}
&&\hfill $220(\tro)$, $100(\tpi)$\hfill&&\hfill$5$\hfill&&
\hfill$35$ \hfill& \cr\cr
\noalign{\hrule}
height4pt&\omit&&\omit&&\omit&&\omit&\cr\cr
&$\tro\ra\tpi\tpi${$\ts ^{1}$} &&\hfill
$220(\tro)$, $100(\tpi)$\hfill&&\hfill$5$\hfill&&
\hfill$25$\hfill&\cr\cr
\noalign{\hrule}
height4pt&\omit&&\omit&&\omit&&\omit&\cr\cr
&$gg \ra \tpiz \ra b \ol b${$\ts ^{2}$}&&\hfill
$100$\hfill&& \hfill$300/5000$\hfill&&\hfill
$7000/10^{5}$\hfill& \cr\cr
\noalign{\hrule}
height4pt&\omit&&\omit&&\omit&&\omit&\cr\cr
&$gg \ra \eta_T \ra t \ol t${$\ts ^{3}$}&&\hfill $400$\hfill&&
\hfill$3/3$\hfill&&\hfill
$2000/600$\hfill& \cr\cr
\noalign{\hrule}
height4pt&\omit&&\omit&&\omit&&\omit&\cr\cr
&$gg\ra\tpi\tpi${$\ts ^{4}$}&&\hfill $100$\hfill&&
\hfill$0.2$\hfill&&\hfill $600$\hfill& \cr\cr
\noalign{\hrule}
height4pt&\omit&&\omit&&\omit&&\omit&\cr\cr
&$\troct\ra \jet\ts\ts\jet${$\ts ^{5}$}&&\hfill $250(\troct)$\hfill&&
\hfill$700/5000$\hfill&&\hfill $1.5\times 10^{4}/1.5\times
10^{5}$\hfill& \cr\cr
%
%
height4pt&\omit&&\omit&&\omit&&\omit&\cr\cr
& &&\hfill $500(\troct)$\hfill&&
\hfill$10/40$\hfill&&\hfill $2000/6000$\hfill& \cr\cr
\noalign{\hrule}
height4pt&\omit&&\omit&&\omit&&\omit&\cr\cr
&$\troct\ra\pi_{T8}\pi_{T8}${$\ts ^{6}$}&&\hfill $550(\troct),
250(\octpi)$\hfill&& \hfill$2$\hfill&&\hfill $2000$\hfill& \cr\cr
\noalign{\hrule}
height4pt&\omit&&\omit&&\omit&&\omit&\cr\cr
&$\troct\ra \tpiql\tpilq${$\ts ^{6}$}&&\hfill$550(\troct),
200(\tpiql)$\hfill&& \hfill$2$\hfill&&\hfill $1000$\hfill& \cr\cr
\noalign{\hrule}
height4pt&\omit&&\omit&&\omit&&\omit&\cr\cr
&$V_8 \ra t \ol t${$\ts ^{7}$}&&\hfill $500$\hfill&&
\hfill$8/3$\hfill&&\hfill $100/600$\hfill& \cr\cr
\noalign{\hrule}
height4pt&\omit&&\omit&&\omit&&\omit&\cr\cr
&$\Lambda$ reach{$\ts ^{8}$}&&\hfill 5~TeV (TeV),
20~TeV (LHC)\hfill&& \hfill$10\,\fb^{-1}$\hfill&&\hfill
$100\,\fb^{-1}$\hfill& \cr\cr 
height4pt&\omit&&\omit&&\omit&&\omit&\cr\cr}
\hrule\hrule}}

\bigskip

\noindent $^{1}$ $F_T = F_\pi/3 = 82\,\gev$ was used.

\noindent $^{2}$ $F_T = 50\,\gev$ used. Cross
section is integrated over $\CM_{b \ol b} = 90$--$110\,\gev$.

\noindent $^{3}$ $F_T = 50\,\gev$ and $m_t = 175\,\gev$ were used. The
greatly increased LHC cross section is due to the rapid growth of gluons at
small-$x$.

\noindent $^{4}$ Cross sections for a multiscale model with 250~GeV
$\octpi$ and 200~GeV $\tpiql$ intermediate states.

\noindent $^{5}$ Jet energy resolution of $\sigma(E)/E = 100\%/\sqrt{E}$
is assumed and cross sections integrated over $\pm \Gamma$ about resonance
peak. Jet angles are limited by $\cos \theta^* < \twothirds$ and $\vert
\eta_j \vert < 2.0$ (Tevatron) or 1.0 (LHC).

\noindent $^{6}$ Cross sections per channel are quoted.

\noindent $^{7}$ $\tan\xi = \sqrt{2\pi/3\aqcd}$ was used, corresponding to a
critical topcolor coupling strength.

\noindent $^{8}$ Estimated $\Lambda$ reaches in dijet and dilepton production
are for the indicated luminosities.


\centerline{\vbox{\offinterlineskip
\hrule\hrule
\halign{&\vrule#&
  \strut\quad#\hfil\quad\cr\cr
height4pt&\omit&&\omit&&\omit&&\omit&\cr\cr
&\hfill Process \hfill&&\hfill
References \hfill&&\hfill CDF Contact \hfill&&\hfill D\O\ Contact \hfill
&\cr\cr
height4pt&\omit&&\omit&&\omit&&\omit&&\omit&&\omit&\cr\cr
\noalign{\hrule\hrule}
height4pt&\omit&&\omit&&\omit&&\omit&\cr\cr
&$\tro\ra W_L\tpi$ &&\hfill \multi, \tctpi\ \hfill&&\hfill
toback@fnald\hfill&&\hfill
hobbs@d0sgi4\hfill&\cr\cr
%
height4pt&\omit&&\omit&&\omit&&\omit&\cr\cr
&  &&\hfill \hfill&&\hfill
\hfill &&\hfill
womersley@fnald0\hfill&\cr\cr
\noalign{\hrule}
height4pt&\omit&&\omit&&\omit&&\omit&\cr\cr
&$\tro\ra\tpi\tpi$ &&\hfill \multi, \tctpi\ \hfill&&\hfill
troconiz@fnald\hfill&&\hfill
womersley@fnald0\hfill&\cr\cr
\noalign{\hrule}
height4pt&\omit&&\omit&&\omit&&\omit&\cr\cr
&$gg \ra \tpiz \ra b \ol b$ &&\hfill \ehlq,\etatrefs,\cdfcoloron\
\hfill&&\hfill
benlloch@fnald\hfill &&\hfill
johns@fnald0\hfill&\cr\cr
%
height4pt&\omit&&\omit&&\omit&&\omit&\cr\cr
&  &&\hfill \hfill&&\hfill
\hfill &&\hfill
zieminski@fnald0\hfill&\cr\cr
\noalign{\hrule}
height4pt&\omit&&\omit&&\omit&&\omit&\cr\cr
&$gg \ra \eta_T \ra t \ol t$ &&\hfill \ehlq,\etatrefs,\cdfcoloron\
\hfill&&\hfill
kirsten@fnald\hfill &&\hfill
klima@fnal\hfill&\cr\cr
\noalign{\hrule}
height4pt&\omit&&\omit&&\omit&&\omit&\cr\cr
&$gg\ra\tpi\tpi$ &&\hfill \tpitev\ \hfill&&\hfill
troconiz@fnald\hfill &&\hfill
womersley@fnald0\hfill&\cr\cr
\noalign{\hrule}
height4pt&\omit&&\omit&&\omit&&\omit&\cr\cr
&$\troct\ra \jet\ts\ts\jet$ &&\hfill \multi, \cdfdijet\ \hfill&&\hfill
rharris@cdfsga\hfill&&\hfill
bertram@fnald0\hfill&\cr\cr
%
height4pt&\omit&&\omit&&\omit&&\omit&\cr\cr
&  &&\hfill \hfill&&\hfill
chaowei@fnald\hfill &&\hfill
\hfill&\cr\cr
\noalign{\hrule}
height4pt&\omit&&\omit&&\omit&&\omit&\cr\cr
&$\troct\ra\pi_{T8}\pi_{T8}$ &&\hfill \multi\ \hfill&&\hfill
troconiz@fnald\hfill &&\hfill
womersley@fnald0\hfill&\cr\cr
\noalign{\hrule}
height4pt&\omit&&\omit&&\omit&&\omit&\cr\cr
&$\troct\ra \tpiql\tpilq$ &&\hfill \multi\ \hfill&& \hfill
baumann@fnald\hfill &&\hfill
womersley@fnald0\hfill&\cr\cr
\noalign{\hrule}
height4pt&\omit&&\omit&&\omit&&\omit&\cr\cr
&$V_8, \ts Z' \ra t \ol t$, $b \ol b$ &&\hfill \hp, \cdfcoloron\ 
\hfill&&\hfill
kirsten@fnald\hfill &&\hfill
klima@fnal\hfill&\cr\cr
%
height4pt&\omit&&\omit&&\omit&&\omit&\cr\cr
&  &&\hfill \hfill&&\hfill
rharris@cdfsga\hfill &&\hfill
womersley@fnald0\hfill&\cr\cr
\noalign{\hrule}
height4pt&\omit&&\omit&&\omit&&\omit&\cr\cr
&$\Lambda$ reach &&\hfill \elp,\ehlq,\cdfjets\ \hfill&&\hfill
rharris@cdfsga (jets)\hfill &&\hfill
hpiekarz@fnald0 \hfill&\cr\cr 
height4pt&\omit&&\omit&&\omit&&\omit&\cr\cr
&  &&\hfill \hfill&&\hfill
\hfill &&\hfill
wightman@fnald0 (jets)\hfill&\cr\cr
%
height4pt&\omit&&\omit&&\omit&&\omit&\cr\cr
&  &&\hfill \gemtdr\ \hfill&&\hfill
maeshima@fnald (leptons)\hfill &&\hfill
eppley@fnald0 (leptons)\hfill&\cr\cr
height4pt&\omit&&\omit&&\omit&&\omit&\cr\cr}
\hrule\hrule}}
\medskip

\bigskip\bigskip

\noindent Table 1. Sample cross sections for technicolor signatures at the
Tevatron and LHC. \hfil\break
Cross sections may vary by a factor of~10 for other masses and choices of
the parameters. $K$-factors of 1.5--2 are expected, but not included.
Signal over background rates are quoted as $S/B$. $\Ntc = 4$ in all
calculations; cross sections generally grow with $\Ntc$. All e-mail
addresses are name@node.fnal.gov unless otherwise noted.


\vfil\eject

\centerline{\vbox{\offinterlineskip
\hrule\hrule
\halign{&\vrule#&
  \strut\quad#\hfil\quad\cr\cr
height4pt&\omit&&\omit&&\omit&&\omit&&\omit&\cr\cr
&\hfill Model \hfill&&\hfill $C_3$ \hfill&&\hfill
$C_8$  \hfill&&\hfill $D_3$\hfill&&\hfill $D_8$\hfill &\cr\cr
height4pt&\omit&&\omit&&\omit&&\omit&&\omit&\cr\cr
\noalign{\hrule\hrule}
height4pt&\omit&&\omit&&\omit&&\omit&&\omit&\cr\cr
&One--Family $\tpi\tpi$&&\hfill${10\over{3}}$\hfill&&\hfill${1\over{3}}$
\hfill&&
\hfill${16\over {9}} M_3^2$\hfill&&\hfill ${4\over{9}}
M_8^2$\hfill&\cr\cr
\noalign{\hrule}
height4pt&\omit&&\omit&&\omit&&\omit&&\omit&\cr\cr
&Multiscale $\pi_{\ol Q Q} \pi_{\ol Q Q}$
&&\hfill${8\over{3}}$\hfill&&\hfill${4\over{3}}$\hfill&&  
\hfill${32\over {9}} M_3^2$\hfill&&\hfill ${16\over{9}}
M_8^2$\hfill&\cr\cr
\noalign{\hrule}
height4pt&\omit&&\omit&&\omit&&\omit&&\omit&\cr\cr
&Multiscale $\pi_{\ol L L} \pi_{\ol L L}$
&&\hfill$8$\hfill&&\hfill$0$\hfill&&  
\hfill${16\over {3}}(2M_{\pi_T}^2 - M_3^2)$\hfill&&\hfill
$0$\hfill&\cr\cr
height4pt&\omit&&\omit&&\omit&&\omit&&\omit&\cr\cr}
\hrule\hrule}}
\medskip

\medskip

\noindent Table 2. The factors $C_R$ and $D_R$ in Eq.~\dsggpp\ for $gg \ra
\tpi\tpi$ for the one--family model and a multiscale technicolor model
containing a doublet of techniquarks $Q$ and technileptons $L$ (see
Ref.~\tpitev). The masses of intermediate color-triplet and octet
technipions are $M_3$ and $M_8$.

\vfil\eject

\bye